\documentclass[journal]{IEEEtran}
\pdfoutput=1
\usepackage{subcaption}
\usepackage{cite}
\usepackage[pdftex]{graphicx}
\DeclareGraphicsExtensions{.pdf,.jpeg,.png}
\usepackage[cmex10]{amsmath}
\usepackage{amssymb}
\usepackage{amsthm}
\usepackage{algorithmic}
\usepackage{array}

\usepackage{mdwmath}
\usepackage{mdwtab}
\usepackage{fixltx2e}
\usepackage{url}

\newtheorem*{theorem*}{Theorem}

\newcounter{customtheoremcounter}
\begin{document}
\title{Hiding Information in Noise: Fundamental Limits of Covert Wireless Communication}

\author{Boulat~A.~Bash,~\IEEEmembership{Student Member,~IEEE,}
        Dennis~Goeckel,~\IEEEmembership{Fellow,~IEEE,}
        Saikat Guha,~\IEEEmembership{Member,~IEEE}
        and~Don~Towsley,~\IEEEmembership{Fellow,~IEEE}
\thanks{Boulat A.~Bash (corresponding author) and Saikat Guha are with Raytheon BBN Technologies.}
\thanks{Dennis Goeckel and Don Towsley are with the University of Massachusetts, Amherst.}
\thanks{This research was sponsored by the National Science Foundation under grants CNS-1018464 and ECCS-1309573.}}
\maketitle
\begin{abstract}
Widely-deployed encryption-based security prevents unauthorized decoding, but 
  does not ensure undetectability of communication.
However, covert, or low probability of detection/intercept (LPD/LPI) 
  communication is crucial in many scenarios ranging from covert military 
  operations and the organization of social unrest, to privacy  protection for
  users of wireless networks.
In addition, encrypted data or even just the transmission of a signal can 
  arouse suspicion, and even the most theoretically robust encryption can
  often be defeated by a determined adversary using non-computational methods 
  such as side-channel analysis.
Various covert communication techniques were developed to address these
  concerns, including steganography for finite-alphabet noiseless applications
  and spread-spectrum systems for wireless communications.
After reviewing these covert communication systems, this 
  article discusses new results on the fundamental limits of their 
  capabilities, as well as provides a vision for the future of such systems.
\end{abstract}

\section{Introduction}
Security and privacy are critical in modern-day wireless communication.
Widely-deployed conventional cryptography presents 
  the adversary with a problem that he/she is assumed not to be able to solve
  because of computational constraints, while
  information-theoretic secrecy presents the adversary
  with a signal from which he/she cannot extract information about the message 
  contained therein.
However, while these approaches address security in many domains by
  protecting the content of the message, they do not mitigate the 
  threat to users' privacy from the discovery of the very existence of the 
  message itself.

Indeed, transmission attempts expose connections between the parties involved,
  and recent disclosures of massive surveillance programs revealed that
  this ``metadata'' is widely collected.
Furthermore, the transmission of encrypted data can arouse 
  suspicion, and many cryptographic schemes can be defeated by a determined 
  adversary using non-computational means such as side-channel analysis.
Anonymous communication tools such as Tor
  resist metadata collection and traffic analysis by randomly directing 
  encrypted messages through a large network.
While these tools conceal the identities of source and destination nodes 
  in a ``crowd'' of relays, they are designed for the Internet and are 
  not effective in wireless networks, which are typically orders of magnitude 
  smaller.
Moreover, such tools offer little protection to users whose communications
  are already being monitored by the adversaries.
Thus, secure communication systems should also 
  provide \emph{covert}, stealth, or low probability of detection/intercept 
  (LPD/LPI) communication.
Such systems not only protect the information contained in the
  message from being decoded, but also prevent the adversary from detecting
  the transmission attempt in the first place and allow communication where
  it is prohibited.

\begin{figure}
\centering
\includegraphics[width=\columnwidth]{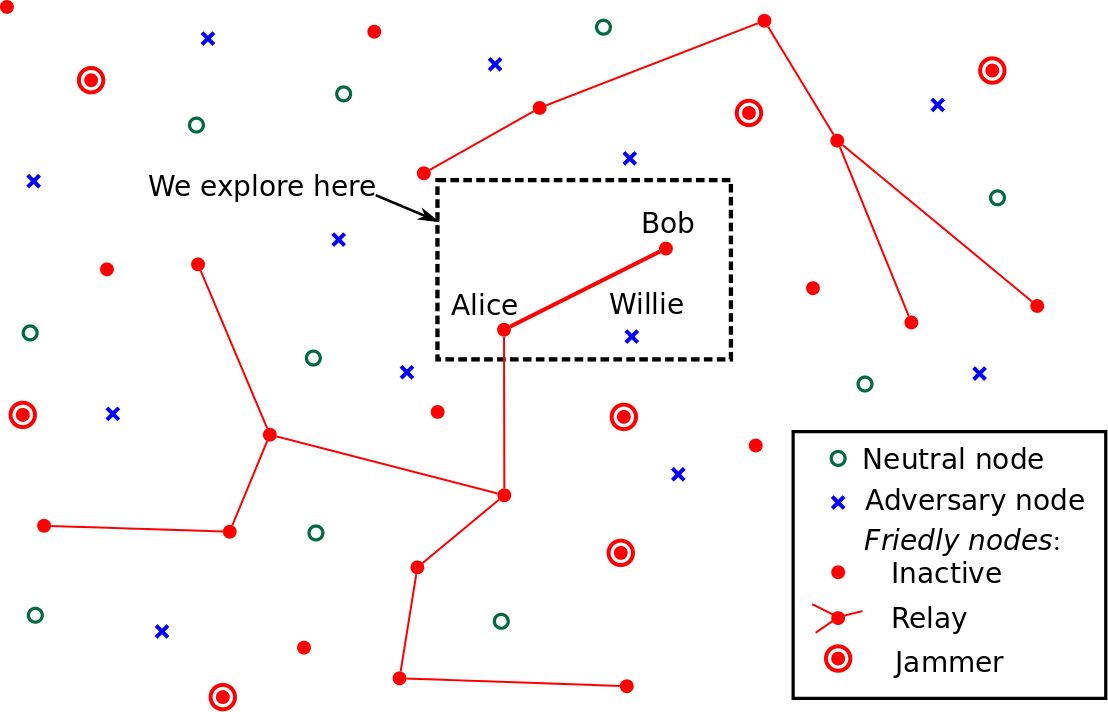}
\caption{Our vision of a ``shadow network''.
  Most of this article focuses on the scenario involving the indicated three 
  nodes: transmitter Alice, receiver Bob, and warden Willie.}
\label{fig:shadownet}
\end{figure}

The overarching goal of covert wireless communication research is
  the establishment of ``shadow networks'' like that depicted 
  in Figure~\ref{fig:shadownet}. 
They are assembled from relays that generate, transmit,
  receive and consume data, and jammers
  that generate artificial noise and impair the ability of wardens to detect
  the presence of communication (we discuss the details of this vision in
  Section~\ref{sec:conclusion}).
However, to create such networks, we must first learn how to connect their
  component nodes by stealthy communication links.
Therefore, in this article we focus on the fundamental limits of 
  such point-to-point links and address the following question: how much 
  information can a sender Alice reliably transmit (if she chooses to 
  transmit) to the intended recipient Bob while hiding it from the adversary, 
  warden Willie?

We begin in Section \ref{sec:stego} by briefly reviewing the field of 
  steganography, or the practice of hiding messages in innocuous objects.
Steganography is important as it was arguably the first covert communication 
  method devised by man.
More recently it has been extensively studied by both the computer science and
  information theory communities in the context of hiding information in 
  digital media.
However, since steganography enables covert communication only at 
  the \emph{application layer}, its analysis has limited use for 
  \emph{physical layer} covert communication techniques such as spread-spectrum.
Therefore, in Section \ref{sec:noisy} we examine the fundamental limits of 
  covert communication over analog radio-frequency (RF) channels, 
  where the information is hidden in the channel artifacts such as additive 
  white Gaussian noise (AWGN), as well as digital communication channels, and 
  briefly touch upon the covert broadcast scenario at the end of the section.
We conclude in Section \ref{sec:conclusion} with a discussion of shadow
  networks and ongoing research in jammer-assisted covert communication.

\section{Steganography}
\label{sec:stego}
Covert communication is an ancient discipline: a description of it is given by
  Herodotus circa 440 BCE in \emph{The Histories}, an account 
  of the Greco-Persian Wars:
        in Chapter 5 Paragraph 35, Histiaeus shaves the head of his slave,
  tattoos the message on his scalp, waits until the hair grows back, and 
  then sends the slave to Aristagoras with instructions to shave the
  head and read the message that calls for an anti-Persian revolt in Ionia; 
  in Chapter 7 Paragraph 239, Demaratus warns Sparta 
  of an imminent Persian invasion by scraping the wax off a wax tablet,
  scribbling a message on the exposed wood, and concealing the message
  by covering the tablet with wax.
This practice of hiding sensitive messages in innocuous objects is known
  as \emph{steganography}.

Modern digital steganography conceals messages in finite-length, 
  finite-alphabet \emph{covertext} objects, such as images or software binary 
  code.
Embedding hidden messages in covertext produces \emph{stegotext}, necessarily
  changing the properties of the covertext.
The countermeasure for steganography, \emph{steganalysis} (an analog of 
  cryptanalysis for cryptography), looks for these changes.
Covertext is usually unavailable for steganalysis (when it is, 
  steganalysis consists of the trivial comparison between the covertext and 
  the suspected stegotext).
However, Willie is assumed to have a complete statistical model of
  the covertext.
The amount of information that can be embedded without being discovered depends
  on whether Alice also has access to this model.
If she does, then \emph{positive-rate steganography} is achievable: given 
  an $\mathcal{O}(n)$-bit\footnote{We use the \emph{Big-O} notation in this
  article, where $\mathcal{O}(f(n))$ denotes an asymptotic upper bound.}
  secret ``key'' that is 
  shared  with Bob prior to the embedding, $\mathcal{O}(n)$ bits can be 
  embedded in an $n$-symbol covertext without being detected by 
  Willie~\cite[Chapter 13.1]{fridrich09stego}.

Recent work focuses on the more general scenario where the complete 
  statistical model of the covertext is unavailable to Alice.
Then, Alice can safely embed $\mathcal{O}(\sqrt{n}\log n)$ bits by modifying 
  $\mathcal{O}(\sqrt{n})$ symbols out of $n$ in the covertext, at the cost
  of pre-sharing $\mathcal{O}(\sqrt{n}\log n)$ secret bits with Bob.
Note that this \emph{square root law of digital steganography} yields 
  \emph{zero-rate steganography} since 
  $\lim_{n\rightarrow\infty}\frac{\mathcal{O}(\sqrt{n}\log n)}{n}=0$ 
  bits/symbol.
The proof is available in Chapter 13.2.1 of the review of pre-2009 work in 
  digital steganography~\cite{fridrich09stego}.
More recent work shows that an \emph{empirical} model of covertext suffices
  to break the square root law and achieve positive-rate 
  steganography~\cite{craver10nosqrtlaw}.
Essentially, while embedding at a positive rate lets Willie obtain
  $\mathcal{O}(n)$ stegotext observations (enabling detection of Alice 
  when statistics of covertext and stegotext differ), the increasing size $n$ 
  of the covertext allows Alice to improve her covertext model and produce
  statistically-matching stegotext.

However, steganography is inherently an application layer covert communication 
  technique.
As such, the results for steganography have limited use in physical layer covert
  communication.
First, analysis of the steganographic systems generally assumes that stegotext
  is not corrupted by a noisy channel.
Second, the generalization of the results for steganographic systems is 
  limited because of their finite-alphabet discrete nature.
Third, by embedding the hidden messages, Alice \emph{replaces} part of 
  the covertext.
While this effectively enables the recent positive rate steganography 
  methods~\cite{craver10nosqrtlaw}, it cannot be done in standard communication
  systems unless Alice controls Willie's noise source.
Finally, the most serious drawback of using steganography for covert 
  communication is the necessity of transmitting the stegotext from 
  Alice to Bob---a potentially unrealizable requirement when all communication 
  is prohibited.
We thus consider physical layer covert communication that employs channel
  artifacts such as noise to hide transmissions.

\section{Physical Layer Covert Communication}
\label{sec:noisy}

\begin{figure*}[ht]
\centering
\begin{subfigure}[b]{.45\linewidth}
\includegraphics[width=\columnwidth]{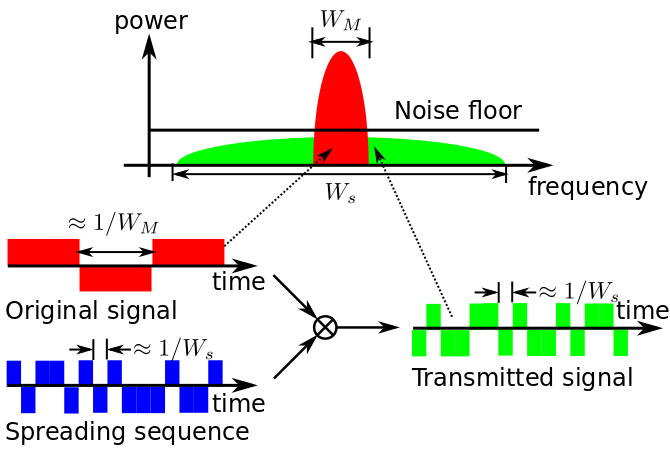}
\caption{DSSS}
\label{fig:dsss}
\end{subfigure}
\qquad
\begin{subfigure}[b]{.45\linewidth}
\includegraphics[width=\columnwidth]{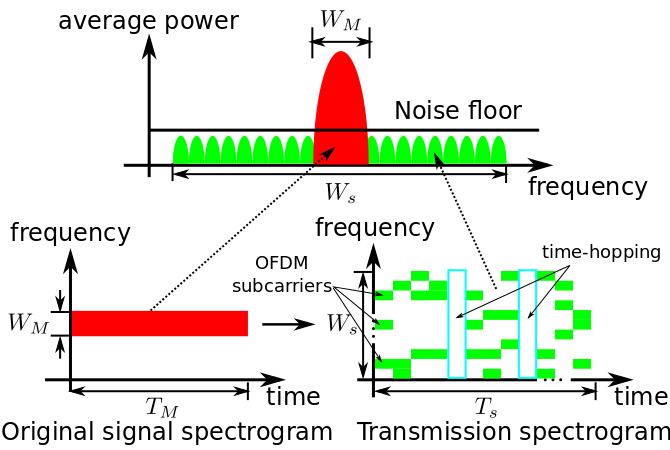}
\caption{FHSS with OFDM and time-hopping.}
\label{fig:fhss}
\end{subfigure}
\caption[Spread spectrum techniques.]{Spread 
  spectrum techniques.}
\label{fig:ss}
\end{figure*}

We begin the investigation of physical layer covert communication 
  by considering RF wireless communication.
Since its emergence in the early 20th century, protecting wireless RF 
  communication from detection, jamming and eavesdropping has been of paramount
  concern.
\emph{Spread spectrum} techniques, devised between the two world wars 
  to address this issue, have constituted the earliest and, arguably, 
  the most enduring form of physical layer security.

\subsection{Spread Spectrum Communication}
\label{sec:ss}
Essentially, the spread spectrum approach involves transmitting a
  signal that requires a bandwidth $W_M$ on a much wider bandwidth 
  $W_s\gg W_M$, thereby suppressing the power spectral density
  of the transmission below the noise floor.
Spread spectrum systems provide both a covert communication capability as well
  as resistance to jamming, fading, and other forms of interference.
A comprehensive review of this field is available in~\cite{simon94ssh}.
Typical spread spectrum techniques include \emph{direct sequence} spread
  spectrum (DSSS), \emph{frequency-hopping} spread spectrum (FHSS), and 
  their combination.

When Alice uses DSSS, she multiplies the signal waveform by 
  the \emph{spreading sequence}---a 
  randomly-generated binary waveform with a substantially higher 
  bandwidth than the original signal. 
The resulting waveform is thus ``spread'' over a wider bandwidth, which 
  reduces the power spectral density of the transmitted signal.
Bob uses the same spreading sequence to de-spread the received
  waveform and obtain the original signal.
The spreading sequence is exchanged by Alice and Bob prior 
  to transmission and is kept secret from 
  Willie.\footnote{While an exchange of a secret prior to covert
  communication is similar to a key exchange in symmetric-key 
  cryptography (e.g., one-time pad), an important distinction is that
  public-key cryptography techniques cannot be used to exchange this secret
  on a channel monitored by Willie without revealing the intention 
  to communicate.}
Outside of security applications, the use of \emph{public} uncorrelated 
  spreading sequences between transmitter/receiver pairs enables multiple 
  access; DSSS thus forms the basis of code-division multiple access (CDMA) 
  protocols used in cellular telephony.
The operation of DSSS is illustrated in Figure \ref{fig:ss}(\subref{fig:dsss}).

When Alice uses FHSS, she re-tunes the carrier frequency for each transmitted 
  symbol.
However, like the spreading sequence in DSSS, the frequency-hopping pattern is 
  also randomly generated and secretly shared between her and Bob
  prior to the transmission.
FHSS can be combined with orthogonal frequency-division multiplexing (OFDM),
  enabling the use of multiple carrier frequencies.
To further reduce the average transmitted symbol power, FHSS can
  be used with \emph{time-hopping} techniques that randomly vary the
  duty cycle (the time-hopping pattern is also secretly pre-shared between 
  Alice and Bob prior to the transmission).
The operation of FHSS with OFDM and time-hopping is illustrated in Figure
  \ref{fig:ss}(\subref{fig:fhss}).

Although spread spectrum architectures are well-developed, the analytical 
  evaluation of covert communication has been sparse.
A.~Hero studied secrecy as well as undetectability \cite{hero03sstc}
  in a multiple-input multiple-output (MIMO) setting, focusing on
  the signal processing aspects.
He recognized that covert communication systems are constrained by average 
  power, and noted the need to explore the fundamental information-theoretic 
  limits in the conclusion of his work.
In fact, knowledge of the limits of any communication system is 
  important, particularly since modern coding techniques (such as 
  Turbo codes and low-density parity check codes) allow 3G/4G cellular 
  systems to operate near their theoretical \emph{channel capacity}, the
  maximum rate of reliable communication that is unconstrained by the security 
  requirements.
However, while the secrecy portion of \cite{hero03sstc} has drawn significant 
  attention, the covert communication portion has been largely overlooked
  until our work on the square root limit of covert communication that we
  discuss next.
We note that the fundamental results that follow apply not
  only to the classical spread-spectrum systems, but also to the modern covert
  communication proposals that rely on channel noise and equipment imperfections
  to hide communications (as is done, in, e.g.,~\cite{dutta13dirtyconstel}).

\subsection{Square Root Law for Covert Communication over AWGN Channels}
\label{sec:awgn}
Spread spectrum systems allow communication where it is prohibited because
  spreading the signal power over a large time-frequency space substantially
  reduces Willie's signal-to-noise ratio (SNR).
This impairs his ability to discriminate between the noise and the 
  information-carrying signal corrupted by noise.
Here we determine just how small the power has to be for the 
  communication to be fundamentally undetectable, and how much covert 
  information can be transmitted reliably.

\begin{figure*}[ht]
\centering
\begin{subfigure}[b]{.30\linewidth}
\includegraphics[scale=0.16]{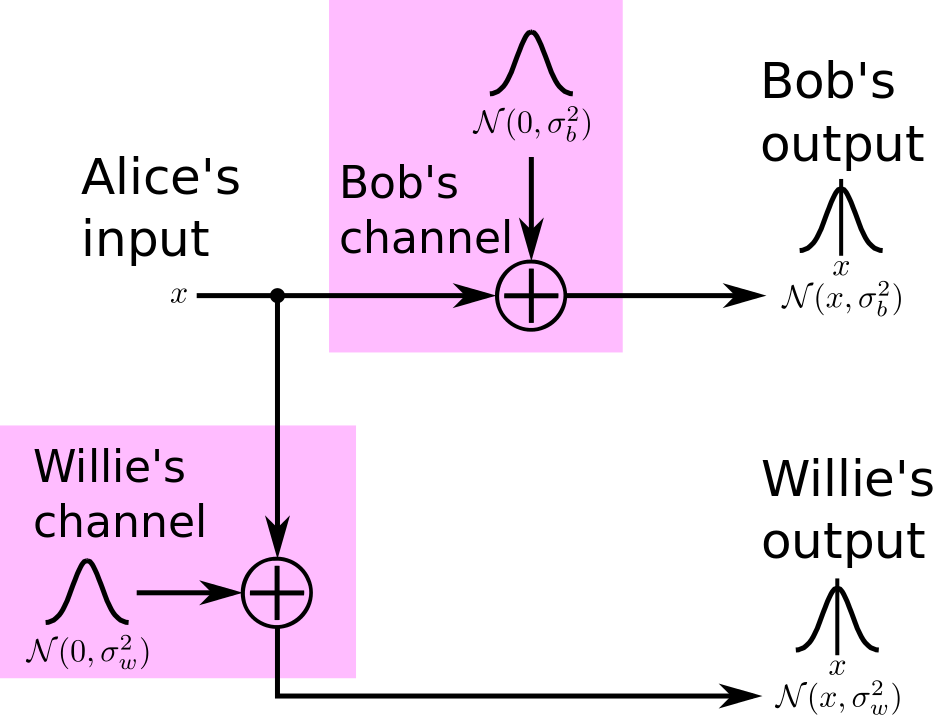}
\caption{AWGN channel}
\label{fig:awgn}
\end{subfigure}
\qquad
\begin{subfigure}[b]{.30\linewidth}
\includegraphics[scale=0.16]{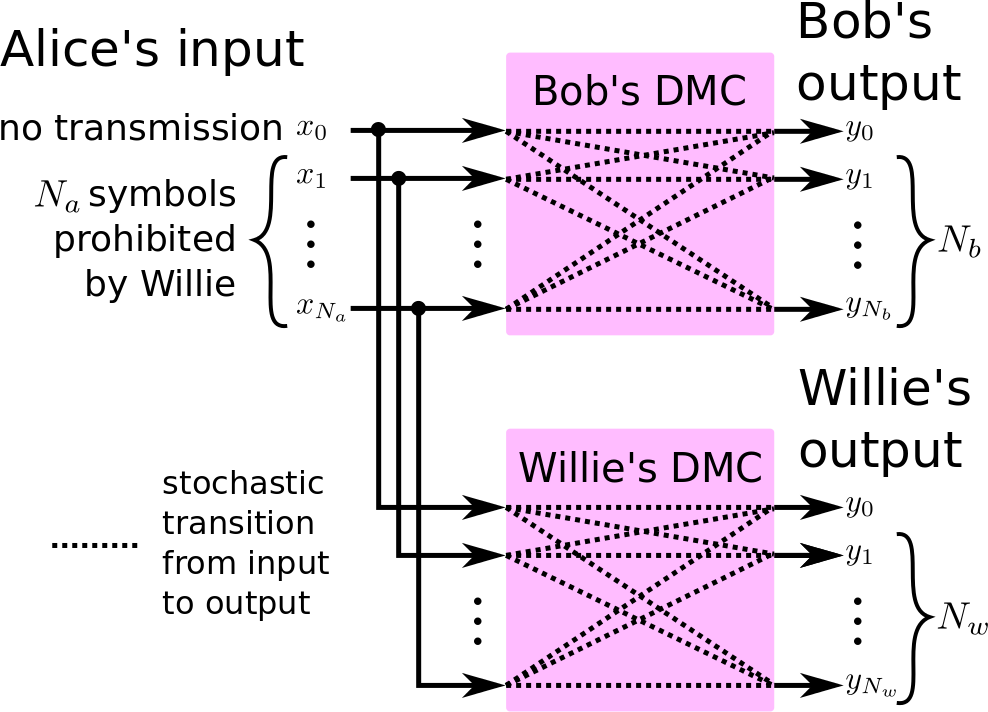}
\caption{DMC}
\label{fig:dmc}
\end{subfigure}
\qquad
\begin{subfigure}[b]{.30\linewidth}
\includegraphics[scale=0.16]{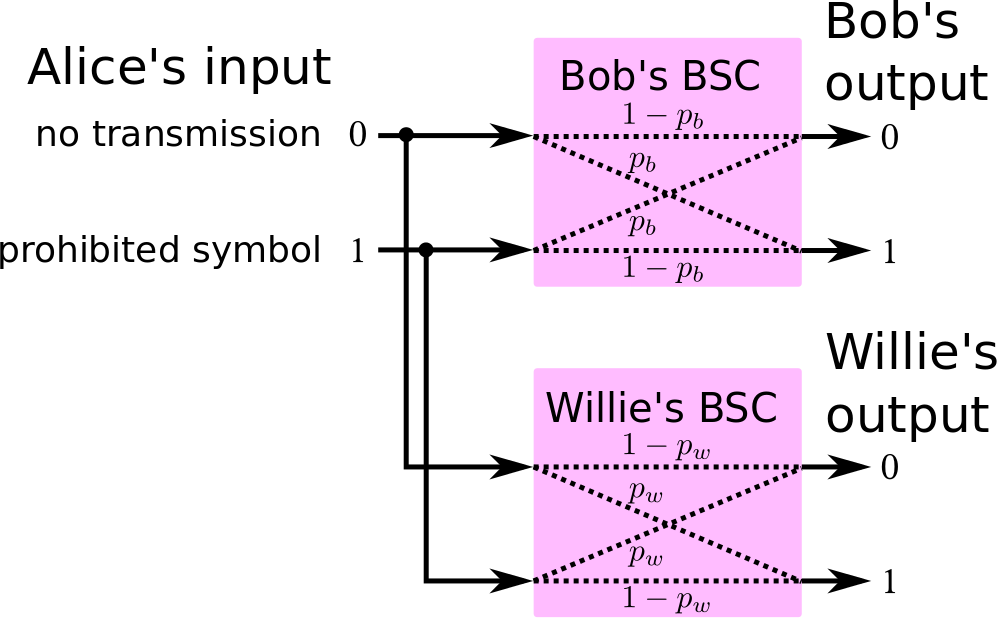}
\caption{BSC}
\label{fig:bsc}
\end{subfigure}
\caption[Channel models.]{Channel models.}
\label{fig:models}
\end{figure*}

Consider an additive white Gaussian noise (AWGN) channel model where 
  the signaling sequence is corrupted by the addition of a sequence 
  of independent and identically distributed zero-mean Gaussian random 
  variables with variance $\sigma^2$.
This is the standard model for a free-space RF channel.
Suppose that the channels from Alice to Bob and to Willie are
  subject to AWGN with respective variances $\sigma_b^2>0$ and
  $\sigma_w^2>0$,\footnote{If the channel from Alice to Bob is noiseless 
  ($\sigma_b^2=0$) and the channel from Alice to Willie is noisy 
  ($\sigma_w^2>0$), then Alice can transmit an infinite amount of information 
  to Bob; if the channel from Alice to Willie is noiseless ($\sigma_w^2=0$),
  then covert communication is impossible.}
  as illustrated in Figure~\ref{fig:models}(\subref{fig:awgn}).
Let \emph{channel use} denote the unit of communication resource---a fixed 
  time period that is used to transmit a fixed-bandwidth signal---and let $n$ 
  be the total number of channel uses available to Alice and Bob (e.g., 
  $n=W_sT_s$ in Figure \ref{fig:ss}(\subref{fig:fhss})).
Willie's ability to detect Alice's transmission depends on the amount of total 
  power that she uses.
Let's intuitively derive\footnote{The formal proof is 
  in~\cite[Section III]{bash13squarerootjsacisit}.} Alice's power constraint 
  assuming that Willie observes these $n$ channel uses.
When Alice is not transmitting, Willie observes AWGN
  with total power $\sigma_w^2n$ over $n$ channel observations on average.
By standard statistical arguments, with high probability, observations of
  the total power lie within $\pm c\sigma_w^2\sqrt{n}$ of this average, where
  $c$ is a constant.
Since Willie observes Alice's signal power when she transmits in addition to 
  the noise power, to prevent Willie from getting suspicious, the total 
  power that Alice can emit over $n$ channel uses is limited to 
  $\mathcal{O}(\sigma_w^2\sqrt{n})$; otherwise her transmission will be
  detected (in fact, a standard radiometer suffices for Willie to detect 
  her if she emits more power, provided $\sigma_w^2$ is 
  known\footnote{See \cite[Section IV]{bash13squarerootjsacisit} for 
  the proof.}). %
This allows her to reliably transmit 
  $\mathcal{O}(\sigma_w^2\sqrt{n}/\sigma_b^2)$ covert bits to Bob in $n$ 
  channel uses, but no more than that~\cite{bash13squarerootjsacisit}.
Note that, just like the steganographic square root law from 
  Section~\ref{sec:stego}, this yields a zero-rate channel 
  (as $\lim_{n\rightarrow\infty}\frac{\mathcal{O}(\sqrt{n})}{n}=0$ bits/symbol).
The similarity of this \emph{square root law for covert communications} to
  the steganographic square root law
  is attributable to the mathematics of statistical hypothesis testing.
The additional $\log n$ factor in the steganographic square root law comes 
  from the fact that the steganographic ``channel'' to Bob is noiseless.

\begin{figure*}
\centering
\includegraphics[scale=0.2]{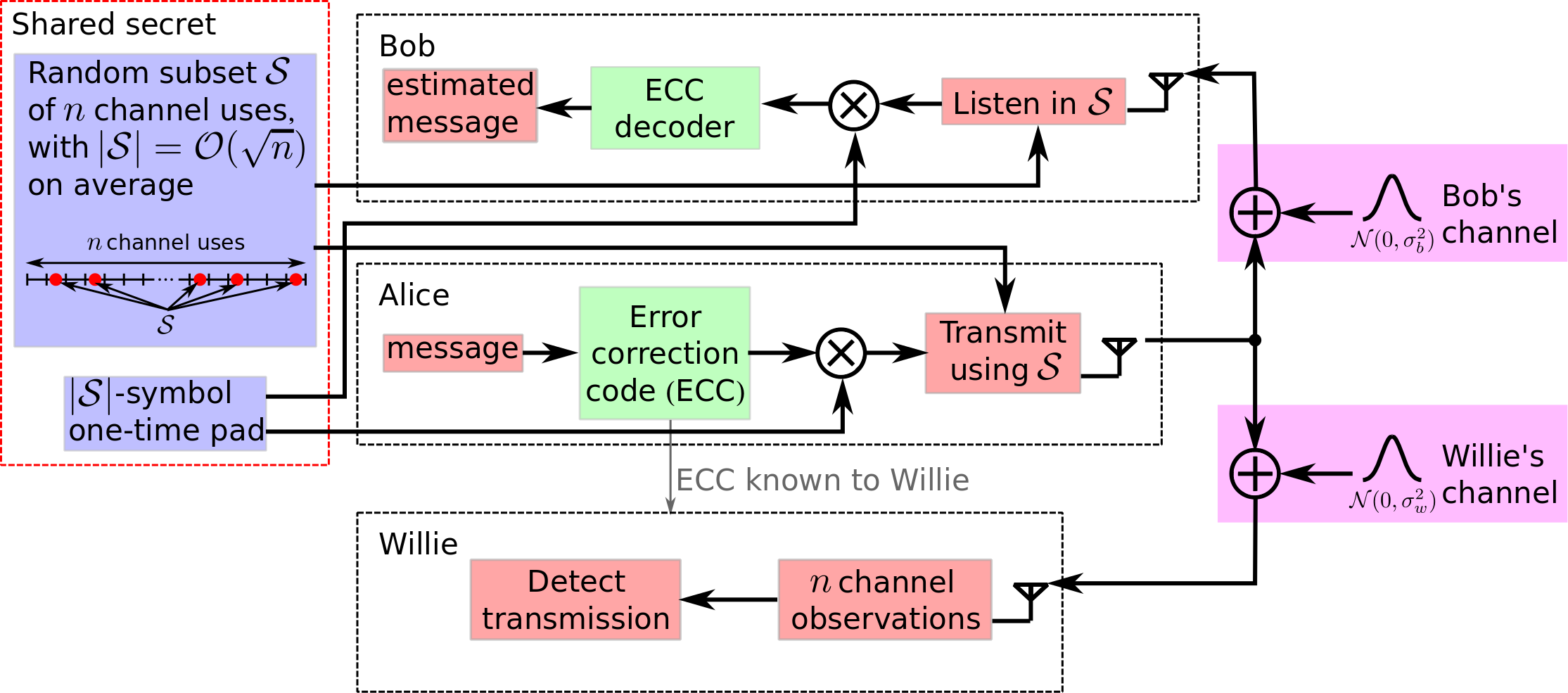}
\caption[Design of a covert communication system.]{Design of a covert 
  communication system that allows Alice and Bob to use
  any error-correction codes (including those known to Willie) 
  to reliably transmit $\mathcal{O}(\sqrt{n})$ 
  covert bits using $\mathcal{O}(\sqrt{n}\log n)$ pre-shared secret 
  bits.}
\label{fig:arbcode}
\end{figure*}

As in steganography and spread spectrum communication, prior to 
  communicating, Alice and Bob may share a secret.
For example, a scheme described in~\cite{bash13squarerootjsacisit} and
  depicted in Figure~\ref{fig:arbcode} allows 
  Alice and Bob to reliably transmit 
  $\mathcal{O}(\sigma_w^2\sqrt{n}/\sigma_b^2)$ covert bits using
  binary amplitude modulation,
  any error-correction code (which can be known to Willie), and
  $\mathcal{O}(\sqrt{n}\log n)$ pre-shared secret bits.
The secret contains a random subset $\mathcal{S}$ of $n$ available 
  channel uses (effectively a frequency/time-hopping pattern),
  and a random one-time pad of size $|\mathcal{S}|$.
$\mathcal{S}$ is generated by flipping a biased random coin $n$ times with
  probability of heads $\mathcal{O}(1/\sqrt{n})$: the $i^{\text{th}}$ channel 
  use is selected for transmission if the $i^{\text{th}}$ flip is heads;
  on average, $|\mathcal{S}|=\mathcal{O}(\sqrt{n})$.
Knowledge of $\mathcal{S}$ allows Bob to discard the observations 
  that are not in $\mathcal{S}$ and decode Alice's message;
  Willie observes mostly noise since he does not have $\mathcal{S}$.
Rather than protecting the message content, the one-time pad prevents Willie's
  exploitation of the error correction code's structure to detect Alice.

While the size of the key is asymptotically larger than the size of the 
  transmitted message, there are many real-world scenarios where this is an
  acceptable trade-off to being detected.
Furthermore, the recent extension of~\cite{bash13squarerootjsacisit} to
  digital covert communication that we describe next suggests that 
  the pre-shared secret can be eliminated in some scenarios.

\subsection{Digital Covert Communication}
\label{sec:dmc}
The \emph{discrete memoryless channel} (DMC) model
  describing digital communication often sheds light on what is feasible 
  in practical communication systems.
DMC model assumes discrete input and output, which allows the DMC to be 
  represented using a bipartite graph where the two sets of vertices correspond
  to input and output alphabets, and edges 
  correspond to the stochastic transitions from input to output symbols.
The memoryless nature of the DMC means that its output is statistically 
  independent from any symbol other than the input at that time.
We illustrate this model in Figure~\ref{fig:models}(\subref{fig:dmc}), which
  we augment by designating one of Alice's inputs as 
  ``no transmission''---a necessary default channel input permitted by 
  Willie.\footnote{For example, this could be the zero-signal in the AWGN 
  channel scenario.}

We first consider the \emph{binary symmetric channel} (BSC) illustrated in
  Figure~\ref{fig:models}(\subref{fig:bsc}), which restricts
  the DMC to binary input and output alphabet $\{0,1\}$, and the probability 
  of a crossover from zero at the input to one at the output 
  being equal to that of a crossover from one to zero.
Denote by $p_b>0$ and $p_w>0$ the crossover probabilities on Bob's and Willie's
  BSCs, respectively.
It has been shown that, while no more than $\mathcal{O}(\sqrt{n})$ 
  covert bits can be reliably transmitted in $n$ BSC uses, if $p_w>p_b$, 
  then the pre-shared secret is unnecessary~\cite{che13sqrtlawbsc}.

Channel \emph{resolvability} can be employed to generalize the square root 
  law in~\cite{che13sqrtlawbsc} to DMCs.
Channel resolvability is the minimum input
  entropy\footnote{Essentially, entropy measures ``surprise'' associated with 
  a random variable, or its ``uncertainty''.  
For example, a binary random variable describing a flip of a fair coin with 
  equal probabilities of heads and tails has higher entropy than the binary
  random variable describing a flip of a biased coin with probability
  of heads larger than tails.
  The output of the biased coin is more predictable, and less surprising,
  as one should observe more heads.
  Introductory texts on the information theory 
  provide the in-depth discussion of entropy and other information-theoretic 
  concepts.} needed
  to generate a channel output distribution that is ``close'' (by some measure
  of closeness between probability distributions\footnote{Examples of measures 
  of closeness are variational distance and relative 
  entropy.}) to 
  the channel output distribution for a given input;
  resolvability has been used to obtain new, stronger results for the 
  information-theoretic secrecy capacity~\cite{bloch13resolvability}.
If the channels from Alice to both Willie and Bob are DMCs, and Willie's
  channel is worse than Bob's, then techniques 
  in~\cite{che13sqrtlawbsc,hou14isit} can be used to demonstrate the square
  root law without a pre-shared secret~\cite{bloch15covert-arxiv}.
Furthermore, as long as the Alice-to-Willie channel is known to Alice,
  $\mathcal{O}(\sqrt{n})$ pre-shared secret bits are 
  sufficient for covert communication when Willie's channel
  capacity is greater-than-or-equal to Bob's~\cite{bloch15covert-arxiv}.
The results in~\cite{bloch15covert-arxiv} can be adapted to AWGN channels as 
  well: a covert communication scheme exists\footnote{Conceptually the covert 
  communication scheme that uses $\mathcal{O}(\sqrt{n})$ secret bits
  resembles the method that uses $\mathcal{O}(\sqrt{n}\log n)$ secret bits
  as described in Figure \ref{fig:arbcode} and Section \ref{sec:awgn}; however, 
  its mathematical analysis is highly technical and is outside the scope of this
  article.} that uses $\mathcal{O}(\sqrt{n})$  pre-shared secret bits, and, 
  if the noise power at Willie's receiver is greater than that at 
  Bob's receiver, then secret-less covert communication is achievable.

\subsection{Willie's Ignorance of Transmission Time Helps Alice}
\label{sec:timing}
When deriving the square root laws, we assume that Willie 
  knows \emph{when} the transmission takes place, if it does.
However, in many practical scenarios Alice and Bob have a pre-arranged time
  for communication that is unknown to Willie (e.g., a certain time and day).
The transmission might also be short relative to the total time during which it
  may take place (e.g., a few seconds out of the day).
If Willie does not know when the message may be transmitted, he has to monitor 
  a much longer time period than the time required for the transmission.
It turns out that Willie's ignorance of Alice's transmission time allows her
  to transmit additional information to Bob.
Surprisingly, under some mild conditions on the relationship between the total
  available transmission time and the transmission duration,
  Alice and Bob do not even have to pre-arrange the communication time.
The technical details of this work are provided in~\cite{bash14timingarxiv}.

\subsection{Positive-rate Covert Communication}
The covert communication channels described above are zero-rate, since the 
  average number of bits that can be covertly transmitted per channel use tends
  to zero as the number of channel uses $n$ gets large.
Here we discuss the possibility of positive-rate covert communication,
  i.e.~reliable transmission of $\mathcal{O}(n)$ covert bits in $n$ channel
  uses.
In general, the circumstances that allow Alice to covertly communicate with Bob
  at positive rates occur either when Willie \emph{allows} Alice to transmit
  messages containing information (rather than zero-signal) or when he is 
  ignorant of the probabilistic structure of the noise on his channel
  (note that the applicability of the steganographic 
  results~\cite{craver10nosqrtlaw} here is limited since estimation of 
  the probabilistic structure of the noise on Willie's channel is insufficient
  unless Alice can ``replace'' this noise rather than add to it).
When Willie allows transmissions, the covert capacity is the same as the
  information-theoretic secrecy capacity (see~\cite{hou14isit} for treatment 
  of the DMCs).
Incompleteness of Willie's noise model can also allow positive-rate covert 
  communication:
  in the noisy digital channel setting, Willie's ignorance of the channel model
  is a special case of the scenario in~\cite{hou14isit}; while in the AWGN 
  channel setting, random noise power fluctuations have been shown to yield
  positive-rate covert communication~\cite{lee15posratecovertjstsp}.
The latter result holds even when the noise power can be bounded; a positive 
  rate is achieved because Willie does not have a constant baseline of noise for
  comparison.

\subsection{Covert Broadcast}
\label{sec:broadcast}
Some of the results for the point-to-point covert communication in the presence
  of a single warden that are discussed in this section can easily be extended
  to scenarios with multiple independently-controlled receivers. %
For example, covert communication over an AWGN channel effectively imposes 
  a power constraint on Alice.
Since a pre-shared secret enables covert communication in this setting, 
  if each receiver obtains it prior to communication, Alice can use
  standard techniques from network information theory to encode
  covert messages to multiple recipients.
The extension to a multi-warden setting as well as other networked
  scenarios is the ongoing work discussed 
  next.

\section{Conclusion: Towards Shadow Networks}
\label{sec:conclusion}
Our ultimate objective is to enable a wireless ``shadow network'', 
  illustrated in Figure \ref{fig:shadownet}, 
  comprised of transmitters, receivers, and friendly jammers that generate
  artificial noise, impairing wardens' ability to detect transmissions.
While the relays are 
  valuable and require protection, the jammers can be cheap, numerous, and 
  disposable (i.e., the adversary can silence a particular jammer easily, but, 
  because of their great numbers, silencing enough of them to produce 
  a significant impact is infeasible).
Thus, jammers have been shown to facilitate information-theoretically secrecy
  by confusing the eavesdropper even while being completely
  ignorant of the messages exchanged by legitimate communicating 
  parties~\cite{lai08jammers}.

In covert networks jammer activities are independent from the relay transmission states: that 
  is, wardens cannot detect transmissions by listening to the jammers.
Thus, jammers have a parasitic effect on the wardens' SNRs and are
  a nuisance.
It is important to characterize the scaling behavior of such a network, akin
  to the recent results for the secure (but not covert) multipath unicast 
  communication in large wireless networks~\cite{capar12secrecy}.
The first step towards this goal is extending 
  the covert communication scenario of this article to point-to-point 
  jammer-assisted covert communication in the presence of multiple 
  wardens.
Preliminary results~\cite{soltani14netlpdallerton} assume that jammers 
  operate at a constant power, and the signal propagation model accounts only 
  for path loss and AWGN.
However, as~\cite{lee15posratecovertjstsp} demonstrates,
  uncertainty in noise experienced by the warden is beneficial to Alice.
Thus, variable jamming power and multipath fading should be incorporated
  into the jammer-assisted covert communication model, as it may enable
  covert communication at a positive rate.
Completing the characterization of the point-to-point covert link in a 
  multi-warden multi-jammer environment is an important step towards 
  understanding the behavior of ``shadow networks'', and their eventual 
  implementation.

\section*{Acknowledgement}
We are grateful to Matthieu Bloch, Gerhard Kramer, and Sid Jaggi  for
  discussions and comments on a draft of this article.

\vspace*{-2.5\baselineskip}
\begin{IEEEbiographynophoto}{Boulat A.~Bash}
holds BA (2001) in Economics from Dartmouth College, and MS (2008) and PhD (2015) in Computer Science from the University of Massachusetts, Amherst. He is currently a Postdoctoral Fellow in the Quantum Information Processing (QuIP) group at Raytheon BBN Technologies. His research interests include security, privacy, communications, signal processing, and information theory.
\end{IEEEbiographynophoto}
\vspace*{-2.5\baselineskip}
\begin{IEEEbiographynophoto}{Dennis Goeckel}
holds BS (1992) from Purdue University, and MS (1993) and PhD (1996) from the University of Michigan.  He is a Professor in the Electrical and Computer Engineering Department at the University of Massachusetts, Amherst.
His research interests include physical layer communications and wireless network theory.
He received the NSF CAREER Award (1999) and is an IEEE Fellow. He was a Lilly Teaching Fellow (2000-2001), and received the University of Massachusetts Distinguished Teaching Award (2007).  

\end{IEEEbiographynophoto}
\vspace*{-2.5\baselineskip}
\begin{IEEEbiographynophoto}{Saikat Guha}
holds BTech in Electrical Engineering (2002) from the Indian Institute of Technology, Kanpur, and SM (2004) and PhD (2008) in Electrical Engineering and Computer Science from the Massachusetts Institute of Technology. He is a Senior Scientist in the Quantum Information Processing (QuIP) group at Raytheon BBN Technologies. His research interests span quantum-optical communication and sensing, optical quantum computing, and network information theory. 
He received a NASA Tech Brief Award in 2010.
\end{IEEEbiographynophoto}
\vspace*{-2.5\baselineskip}
\begin{IEEEbiographynophoto}{Don Towsley}
holds BA (1971) in Physics and 
PhD (1975) in Computer Science from University of Texas.  He is
a Distinguished Professor at the University of Massachusetts
in the School of Computer Science. His research interests include
networks and network science.
He received several achievement awards including the 2007 IEEE Koji Kobayashi Award and numerous paper awards. He is a Fellow of both the ACM and IEEE.
\end{IEEEbiographynophoto}
\end{document}